\documentclass{article}

\usepackage[margin=2.5cm]{geometry}

\usepackage{tikz}
\usetikzlibrary{arrows.meta, positioning, shapes.geometric}

\usepackage{amsmath,amssymb,amsfonts,mathtools,mathrsfs,bm}
\usepackage{amsthm} 
\usepackage{tabularx,booktabs,array,longtable,multirow,threeparttable}
\usepackage{graphicx,epsfig,subfigure}
\usepackage{color,colortbl} 
\usepackage{caption}
\usepackage{rotating}
\usepackage{diagbox}
\usepackage{pdflscape,lscape}
\usepackage{float}
\usepackage[dvipsnames]{xcolor}

\usepackage{cite,url,hyperref}
\hypersetup{
  colorlinks=true,
  linkcolor=lightblue,
  urlcolor=lightblue,
  citecolor=lightblue
}

\usepackage{algorithm,algpseudocode}

\usepackage{pifont}
\usepackage{tcolorbox}
\tcbuselibrary{skins}
\usepackage{multicol}

\definecolor{lightblue}{RGB}{70,130,180}   

\newcommand{\RMSection}[2]{\textbf{\hyperref[#2]{#1}}\par}
\newcommand{\RMSub}[2]{\hspace*{1.5em}{\color{gray}\footnotesize$\triangleright$}~\hyperref[#2]{#1}\par}

\theoremstyle{plain}

\theoremstyle{definition}

\theoremstyle{remark}

\title{Partition-of-Unity Gaussian Kolmogorov--Arnold Networks}

\author{
Amir Noorizadegan \\[4pt]
{\small Department of Mathematics, Hong Kong Baptist University, Hong Kong SAR, China} \\
{\small  \texttt{amir\_noori@hkbu.edu.hk}}
}

\begin{document}
\maketitle
\noindent
\begin{abstract}
Gaussian basis functions provide an efficient and flexible alternative to spline activations in Kolmogorov--Arnold networks (KANs), but their performance is sensitive to the Gaussian scale parameter \(\epsilon\). We introduce the partition-of-unity Gaussian KAN (PU-GKAN), a kernel-based normalization framework that transfers the Shepard partition-of-unity principle to trainable KAN edge functions. Each Gaussian edge basis is normalized by its local sum over fixed centers, preserving the standard edge-based architecture and essentially the same trainable coefficient structure as GKAN. The resulting feature map reproduces constants exactly at the edge level and admits an explicit normalized finite-feature kernel interpretation.

Numerical experiments show that PU-GKAN reduces sensitivity to \(\epsilon\), improves validation accuracy for most smooth and moderately non-smooth targets, and provides more stable training behavior. These benefits persist across variations in sample size and number of centers, higher-dimensional architectures, and Matérn RBF bases. PU-GKAN also improves physics-informed neural networks for Helmholtz and wave equations, with validation-error reductions of up to \(90\%\). These results establish partition-of-unity normalization as a mathematically grounded and broadly applicable framework for improving the reliability and accuracy of radial basis function KANs. The implementation is available at \url{https://github.com/AmirNoori68/PU-GKAN}.
\end{abstract}


\section{Introduction}

Kolmogorov--Arnold Networks (KANs) have emerged as a structured alternative to classical multilayer perceptrons (MLPs), replacing fixed node activations with learnable univariate functions placed on edges~\cite{Liu24,Liu24b}. Although MLPs have shown strong potential in scientific computing and AI~\cite{Raissi19,Karniadakis21,Amir24}, they can face challenges in training, accuracy, stability, and generalization for complex problems~\cite{Amir26_reg,Amir24a}. KANs offer a different inductive bias for multivariate approximation, particularly in regression, scientific machine learning, and physics-informed modeling.

The original KAN architecture used B-spline bases~\cite{Liu24}, but many other basis families have since been explored~\cite{Amir_KAN}. These include Chebyshev bases~\cite{SS24,Toscano24_kkan}, Jacobi and rational Jacobi bases~\cite{Aghaei24_fkan,Kashefi25}, Fourier, ReLU, and adaptive piecewise bases~\cite{pde_Rigas24f}, and wavelet, finite-basis, and polynomial formulations~\cite{Bozorgasl24,pde_fbkan_Howard24,Seydi24a}. Related developments include geometric and grid-refinement strategies~\cite{Actor25}, grid-adaptive physics-informed KANs~\cite{pde_Rigas24f}, sparse-identification frameworks~\cite{Howard_2026}, JAX/GPU implementations~\cite{pde_Rigas24f}, and improved initialization methods~\cite{Rigas25_init}. These developments establish the choice of basis as a central design decision in KANs.

Within this landscape, Gaussian radial basis functions (RBFs) are attractive because of their smoothness, locality, and simple analytical form. Beginning with Hardy's multiquadric interpolation~\cite{Hardy71}, RBF methods have developed into a mature framework for interpolation, approximation, and scientific computing. Gaussian KANs therefore connect modern neural architectures with classical kernel and meshfree approximation~\cite{Schaback95,Schaback23,Larsson24}.

Gaussian bases were introduced into KANs through FastKAN~\cite{Li24} as efficient surrogates for spline activations. Subsequent studies developed residual architectures~\cite{Koenig25}, operator-learning formulations with learnable Gaussian bases~\cite{Abueidda25}, methods targeting improved accuracy and training efficiency~\cite{Athanasios2024}, and complex-valued Gaussian KANs~\cite{Wolff25,Che25_complexKAN}. Gaussian-based KANs have also shown promising classification performance, including reduced parameter counts and faster training in high-dimensional or large-scale settings~\cite{Chao26}. These results show that Gaussian bases provide a flexible basis family beyond their original role as spline surrogates.

Despite these advantages, Gaussian KANs retain the classical RBF sensitivity to the scale parameter. Small scales produce overly localized bases, whereas large scales produce nearly flat and ill-conditioned features~\cite{Amir_GKAN}. Conditioning provides a useful diagnostic of this accuracy--stability trade-off in classical RBF methods~\cite{Amir22,Amir23,AmirQC,Amir_eval}. For Gaussian KANs, the resulting analysis gives
\(\epsilon=2/(G-1)\), where \(G\) is the number of Gaussian centers, as a practical upper reference choice~\cite{Amir_GKAN}. This choice often gives accuracy close to that obtained from complete \(\epsilon\)-sweeps in both function approximation and differential-equation problems.

A classical approach to improving RBF stability and accuracy is Shepard-type normalization~\cite{Fasshauer07}. Shepard's method~\cite{Shepard68} normalizes local weights by their sum, producing a partition of unity, and can be interpreted as the zero-degree case of moving least squares. Partition-of-unity methods extend this principle by blending higher-order local approximations through normalized weights~\cite{Franke77}. These ideas suggest a natural normalization of Gaussian KAN bases that preserves locality while avoiding direct use of unnormalized Gaussian responses. The resulting construction is grounded in Shepard approximation, partition-of-unity methods, finite-feature kernels, and conditioning analysis.

The present construction differs from domain-decomposition partition-of-unity methods. Finite-Basis KANs with B-spline bases~\cite{pde_fbkan_Howard24} use overlapping subdomains and smooth window functions, following finite-basis PINNs~\cite{Moseley23,Dolean24} and classical meshfree RBF partition-of-unity methods~\cite{Roberto1,Roberto3,Raeisi_2024a,Raeisi_2024b,Raeisi_2024c,Fardi_2026a,Fardi_2026b}. In these methods, the partition of unity combines several local models. Here, normalization is instead applied directly to the Gaussian basis functions within each KAN edge, without subdomains or local subnetworks.

Motivated by this connection, we introduce the partition-of-unity Gaussian KAN (PU-GKAN), a Shepard-type normalized Gaussian KAN in which the Gaussian basis values are divided by their local sum over fixed centers before applying the learned linear combination. This preserves the standard edge-based KAN architecture and essentially the same trainable coefficient structure as GKAN. Our goal is to investigate the effects of this normalization on robustness and practical accuracy and relate them to the numerical behavior of Gaussian basis expansions.

The main contributions are summarized as follows.

\begin{itemize}
    \item We introduce PU-GKAN, a Shepard-type normalization of Gaussian edge bases over fixed centers. It preserves the standard KAN architecture and provides exact constant reproduction at the edge level.

    \item We present GKAN and PU-GKAN from a finite-feature and additive-kernel viewpoint and adopt the practical scale interval
    \(
    \epsilon\in[1/(G-1),\,2/(G-1)]
    \),
    based on adjacent-center overlap and first-layer conditioning. The upper endpoint often gives accuracy close to the best result from complete \(\epsilon\)-sweeps.

    \item Numerical experiments across different sample sizes, numbers of centers, and higher-dimensional architectures show that PU-GKAN produces more compact internal coordinates and improves accuracy for most tested regression, Helmholtz, and wave problems.

    \item We extend the same partition-of-unity normalization to Matérn RBF bases, showing that the approach is not limited to Gaussian functions.
\end{itemize}

The remainder of the paper is organized as follows. Section~\ref{sec:finite_feature_kernel} presents the finite-feature formulation and PU-GKAN normalization. Section~\ref{subsec:reference_epsilon_range} recalls the reference scale range. Section~\ref{Examples} presents the numerical experiments, and Section~\ref{sec:conclusion} summarizes the findings and possible extensions.

\section{Gaussian KAN from a Finite-Feature and Kernel Viewpoint}
\label{sec:finite_feature_kernel}

Consider layer \(\ell\), mapping
\(x^{(\ell)}\in\mathbb{R}^{n_\ell}\) to
\(x^{(\ell+1)}\in\mathbb{R}^{n_{\ell+1}}\), where \(n_\ell\) is the width of layer \(\ell\). Each edge function is expanded in a finite dictionary
\[
b^{(\ell)}(t)
=
\bigl[
b^{(\ell)}_1(t),\dots,b^{(\ell)}_G(t)
\bigr]^\top
\in\mathbb{R}^{G},
\]
shared by all edges in that layer. The layer output is
\begin{equation}
x^{(\ell+1)}_j
=
\sum_{i=1}^{n_\ell}
\psi^{(\ell)}_{j,i}\!\left(x^{(\ell)}_i\right)
=
\sum_{i=1}^{n_\ell}
\bigl(w^{(\ell)}_{j,i}\bigr)^\top
b^{(\ell)}\!\left(x^{(\ell)}_i\right),
\qquad
j=1,\dots,n_{\ell+1},
\label{eq:kan_layer_general_basis}
\end{equation}
where
\(w^{(\ell)}_{j,i}\in\mathbb{R}^{G}\) contains the trainable coefficients on the edge from input \(i\) to output \(j\).

For \(z=(z_1,\dots,z_{n_\ell})^\top\in\mathbb{R}^{n_\ell}\), define the stacked feature vector
\[
B^{(\ell)}(z)
=
\bigl[
b^{(\ell)}(z_1)^\top,\dots,
b^{(\ell)}(z_{n_\ell})^\top
\bigr]^\top
\in\mathbb{R}^{n_\ell G}.
\]
The feature map induces the scalar and additive layer kernels
\begin{equation}
k_\ell(s,t)
=
\bigl(b^{(\ell)}(s)\bigr)^\top b^{(\ell)}(t),
\qquad
K_\ell(z,z')
=
\bigl(B^{(\ell)}(z)\bigr)^\top B^{(\ell)}(z')
=
\sum_{i=1}^{n_\ell}k_\ell(z_i,z_i'),
\label{eq:layer_kernel_additive_general}
\end{equation}
where \(s,t\in\mathbb{R}\) and
\(z,z'\in\mathbb{R}^{n_\ell}\). Thus, each basis-expanded KAN layer has an additive finite-feature kernel structure. A deep KAN is obtained by composing such layers.

For the Gaussian KAN considered here, we fix \(G\) centers
\(\mathcal{C}=\{c_1,\dots,c_G\}\subset[0,1]\) and a shared scale parameter \(\epsilon>0\). The first-layer inputs are scaled to \([0,1]\), while the hidden-layer coordinates may lie outside this interval. The Gaussian feature map is therefore defined on \(\mathbb{R}\) by
\begin{equation}
\varphi(t)
=
\begin{bmatrix}
\exp\!\left(-\dfrac{(t-c_1)^2}{\epsilon^2}\right)\\
\vdots\\
\exp\!\left(-\dfrac{(t-c_G)^2}{\epsilon^2}\right)
\end{bmatrix}
\in\mathbb{R}^{G}.
\label{eq:gaussian_feature_map_clean}
\end{equation}
The GKAN is obtained from \eqref{eq:kan_layer_general_basis} by taking
\(b^{(\ell)}=\varphi\) in every layer; further details of this formulation are given in~\cite{Amir_GKAN}.

The resulting scalar kernel is
\begin{equation}
k_\epsilon(s,t)
=
\varphi(s)^\top\varphi(t)
=
\sum_{g=1}^{G}
\exp\!\left(-\dfrac{(s-c_g)^2}{\epsilon^2}\right)
\exp\!\left(-\dfrac{(t-c_g)^2}{\epsilon^2}\right).
\label{eq:induced_gaussian_feature_kernel_clean}
\end{equation}
This is the finite-dimensional kernel induced by the fixed-center Gaussian features. It is generally different from the classical translation-invariant Gaussian kernel
\(
\exp(-(s-t)^2/\epsilon^2)
\).

For a sample set
\(
\mathcal{X}=\{x^1,\dots,x^N\}\subset[0,1]^d
\),
where
\(x^n=(x_1^n,\dots,x_d^n)^\top\) and \(d=n_0\), define
\begin{equation}
\Phi^{(i)}
=
\begin{bmatrix}
\varphi(x_i^1)^\top\\
\vdots\\
\varphi(x_i^N)^\top
\end{bmatrix}
\in\mathbb{R}^{N\times G},
\qquad
\Phi
=
\begin{bmatrix}
\Phi^{(1)} & \cdots & \Phi^{(d)}
\end{bmatrix}
\in\mathbb{R}^{N\times dG}.
\label{eq:full_first_layer_matrix_clean}
\end{equation}
Here, \(\Phi^{(i)}\) is the feature matrix associated with input coordinate \(i\), and \(\Phi\) is the full first-layer feature matrix. This matrix will be used later to study the numerical rank and conditioning of the first layer.

\subsection{Shepard-Type Partition-of-Unity Normalization in Gaussian KANs}
\label{subsec:shepard_pu_gkan}

A partition-of-unity Gaussian KAN (PU-GKAN) is obtained by normalizing the Gaussian basis values over the fixed centers. The construction follows the normalization principle of Shepard’s method, in which local weights are divided by their sum~\cite{Shepard68,Fasshauer07}.

For sites \(\{x^1,\dots,x^N\}\), the classical Shepard approximant is
\begin{equation}
P_f(x)
=
\sum_{n=1}^{N}f(x^n)\Psi_n(x),
\qquad
\Psi_n(x)
=
\frac{\omega(x^n,x)}
{\sum_{m=1}^{N}\omega(x^m,x)},
\qquad
\sum_{n=1}^{N}\Psi_n(x)=1,
\label{eq:shepard_classical}
\end{equation}
where \(f(x^n)\) are the sampled values and
\(\omega(x^n,x)\) are nonnegative local weights.

In PU-GKAN, this normalization is applied to the Gaussian basis rather than to data-site weights. Let
\[
\phi_g(t)
=
\exp\!\left(-\frac{(t-c_g)^2}{\epsilon^2}\right),
\qquad
S_\epsilon(t)=\sum_{h=1}^{G}\phi_h(t).
\]
Since \(S_\epsilon(t)>0\) for every \(t\in\mathbb{R}\), the normalized basis is
\begin{equation}
\widetilde{\phi}_g(t)
=
\frac{\phi_g(t)}{S_\epsilon(t)}
=
\frac{\exp\!\left(-\frac{(t-c_g)^2}{\epsilon^2}\right)}
{\sum_{h=1}^{G}
\exp\!\left(-\frac{(t-c_h)^2}{\epsilon^2}\right)},
\qquad
g=1,\dots,G.
\label{eq:normalized_gaussian_basis_component}
\end{equation}
Writing
\(
\widetilde{\varphi}(t)
=
[\widetilde{\phi}_1(t),\dots,\widetilde{\phi}_G(t)]^\top
\in\mathbb{R}^{G}
\),
we have
\begin{equation}
\sum_{g=1}^{G}\widetilde{\phi}_g(t)=1,
\qquad
t\in\mathbb{R}.
\label{eq:normalized_basis_pou}
\end{equation}
PU-GKAN is obtained simply by replacing
\(\varphi\) with \(\widetilde{\varphi}\) in
\eqref{eq:kan_layer_general_basis}. Hence, the original KAN architecture and number of trainable coefficients remain unchanged.

To examine the effect of normalization on the hidden coordinates, we compare GKAN and PU-GKAN for target function \eqref{F1}. Both models use architecture \((2,12,12,1)\), \(G=20\), and fixed scale \(\epsilon=0.1\). The coefficients are trained for \(5000\) epochs using AdamW with learning rate \(10^{-3}\). We use \(N=300\) Halton training points and \(7000\) Halton validation points without added noise.

\begin{figure}[hbt!]
\centering
\includegraphics[width=6.5in]{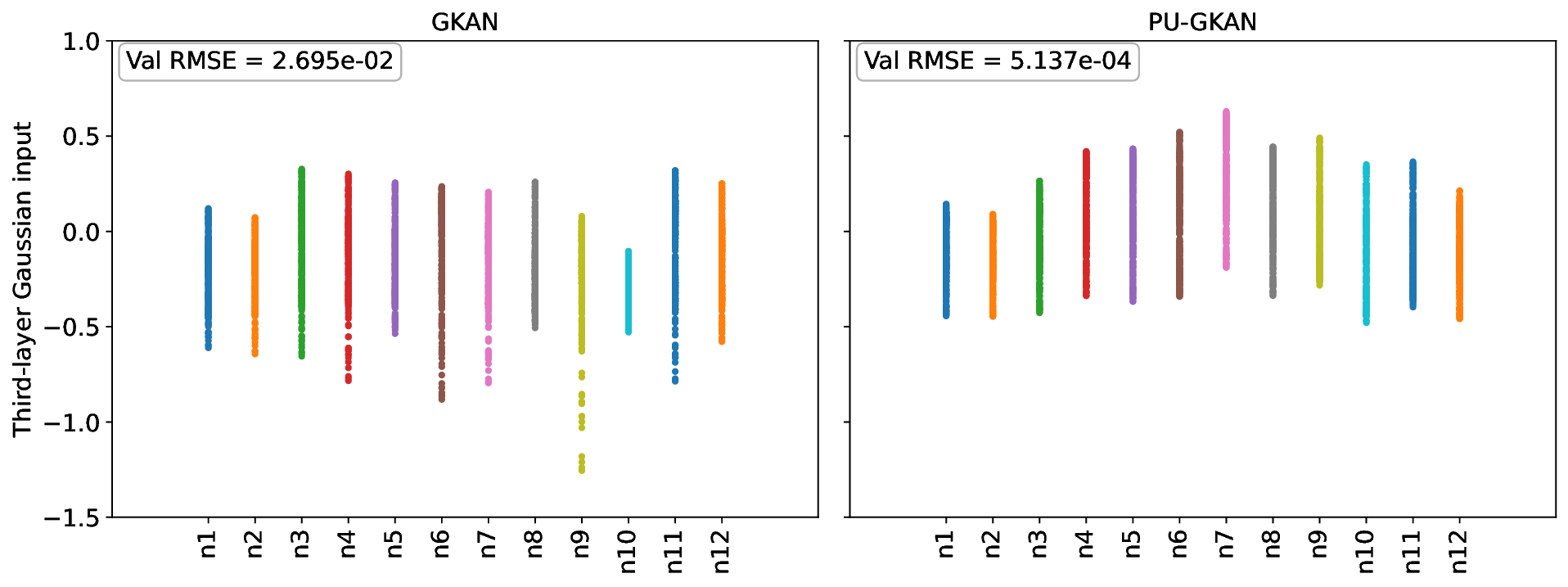}
\caption{Empirical distributions of the third-layer inputs
\(x_i^{(2)}\), labelled \(n_1,\dots,n_{12}\), in GKAN and PU-GKAN.
PU-GKAN produces a more compact distribution with fewer extreme
excursions and attains a lower validation error.}
\label{fig:L3_input_distribution}
\end{figure}

For an input sample \(x^n\), let
\(x^{(2)}(x^n)\in\mathbb{R}^{12}\) denote the input to the third layer. Figure~\ref{fig:L3_input_distribution} shows the samples
\(
\{x_i^{(2)}(x^n)\}_{n=1}^{N}
\)
for coordinates \(i=1,\dots,12\), labelled \(n_1,\dots,n_{12}\) in the figure.

Because the Gaussian centers lie in \([0,1]\), the use of the basis depends on the distribution of the hidden coordinates relative to this interval. Inputs that are strongly clustered or far from the center region activate only a small subset of the basis functions. As shown in Fig.~\ref{fig:L3_input_distribution}, PU-GKAN produces a more compact distribution of third-layer inputs with fewer extreme excursions than GKAN. This behavior is consistent with the lower validation error obtained by PU-GKAN.

To examine the evolution of the trainable coefficients, we record the coefficient norm of each layer after every AdamW update. For layer \(\ell\), the Frobenius norm at epoch \(k\) is
\begin{equation}
\left\|W^{(\ell)}(k)\right\|_{\mathrm{F}}
=
\left[
\sum_{i=1}^{n_\ell}
\sum_{j=1}^{n_{\ell+1}}
\sum_{g=1}^{G}
\left(w^{(\ell)}_{j,i,g}(k)\right)^2
\right]^{1/2},
\label{eq:layer_coefficient_norm}
\end{equation}
where \(w^{(\ell)}_{j,i,g}(k)\) is the coefficient of the \(g\)-th Gaussian basis on the edge from input \(i\) to output \(j\) at epoch \(k\).

\begin{figure}[hbt!]
\centering
\includegraphics[width=6.5in]{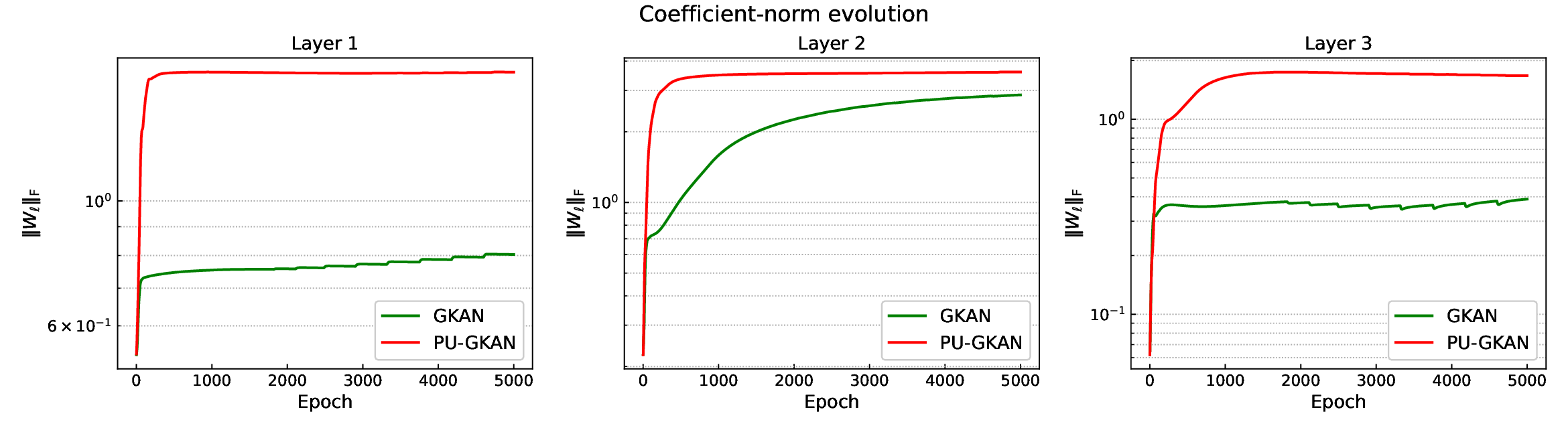}
\caption{Evolution of the layer-wise coefficient Frobenius norms for GKAN and PU-GKAN over \(5000\) epochs. Normalization changes the coefficient scale, while the PU-GKAN norms approach nearly stationary levels in all three layers.}
\label{fig:weight_norm_evolution}
\end{figure}

Figure~\ref{fig:weight_norm_evolution} shows that Shepard normalization changes the numerical scale of the learned coefficients. In PU-GKAN, the coefficient norms increase rapidly during the initial optimization stage and then approach bounded, nearly stationary levels in all three layers. No coefficient explosion is observed. By comparison, GKAN retains smaller norms in the first and third layers, while its second-layer norm increases more gradually throughout training. Since normalization rescales the basis values, the absolute coefficient norms of the two models are not directly comparable as measures of accuracy or conditioning. The relevant observation is that the PU-GKAN coefficients settle rapidly and remain bounded, indicating stable coefficient evolution under the normalized feature representation.

\section{Reference Scale Range for the Gaussian Scale Parameter}
\label{subsec:reference_epsilon_range}

The Gaussian scale parameter \(\epsilon\) controls the overlap of the Gaussian basis functions and strongly affects approximation accuracy and numerical conditioning. In this work, we use the practical scale range derived in our companion study~\cite{Amir_GKAN}, rather than repeating its full derivation.

Let \(\Phi_\epsilon\) denote the first-layer feature matrix \(\Phi\) defined in \eqref{eq:full_first_layer_matrix_clean}, evaluated at scale \(\epsilon\). The first layer acts directly on the original inputs scaled to \([0,1]^d\), with fixed centers in the same interval. Its condition number \(\kappa(\Phi_\epsilon)\) therefore provides a direct measure of the effect of \(\epsilon\). By contrast, the conditioning of deeper layers also depends on the learned hidden coordinates and training dynamics.

\begin{figure}[hbt!]
\centering
\includegraphics[width=6.5in]{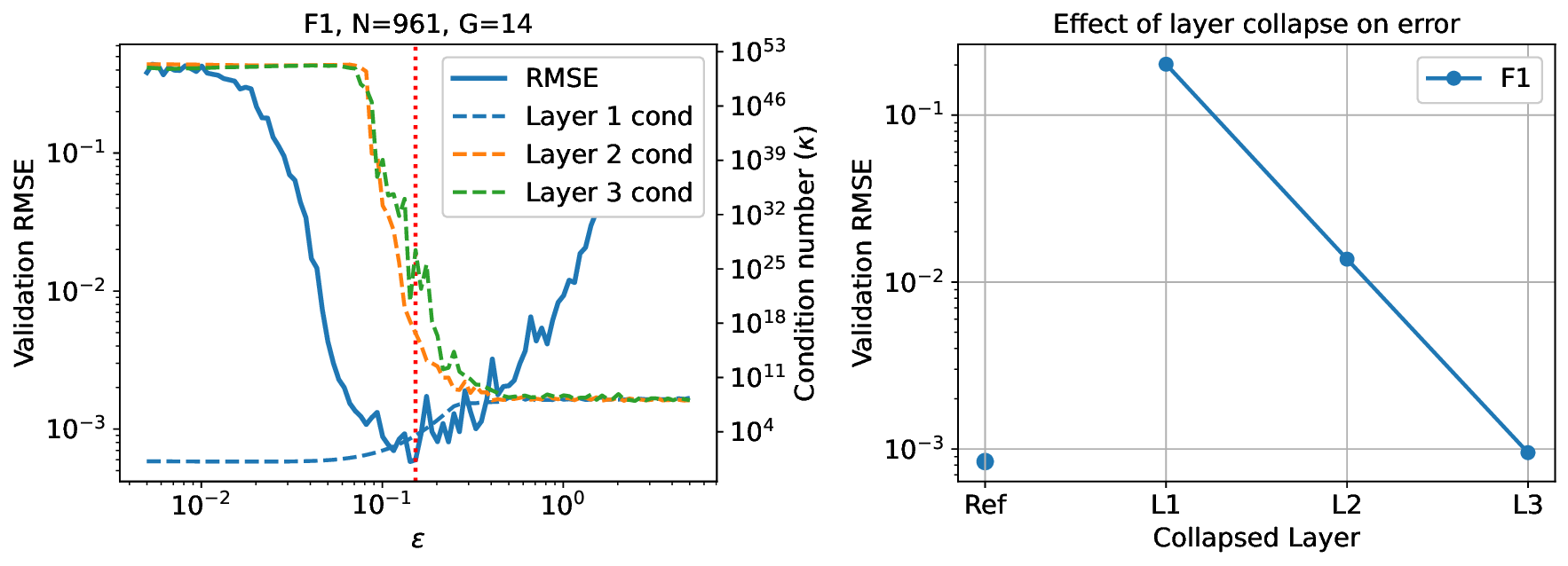}
\caption{Importance of the first layer for \(F1\) with \(N=961\) and \(G=14\). The left panel compares validation RMSE with the layer-wise condition numbers; the red dotted line marks the first-layer conditioning boundary used in~\cite{Amir_GKAN}. The right panel shows that collapsing the first layer causes the largest error increase.}
\label{F1_N961_G14_rmse_cond_with_collapse}
\end{figure}

Figure~\ref{F1_N961_G14_rmse_cond_with_collapse} illustrates the importance of the first layer. The left panel shows that the validation error follows the first-layer conditioning trend more closely than those of the deeper layers. The right panel shows that collapsing the first layer produces the largest increase in error. These observations support using the first-layer feature matrix as the reference for scale selection. The complete conditioning analysis and layer-collapse procedure are presented in~\cite{Amir_GKAN}.

For \(G\) uniformly spaced centers in \([0,1]\), the adjacent-center spacing is \(1/(G-1)\). In~\cite{Amir_GKAN}, this spacing gives the lower reference scale through adjacent-center overlap, while the first-layer conditioning boundary gives the practical upper scale \(2/(G-1)\). We therefore use
\begin{equation}
\boxed{
\epsilon\in
\left[
\frac{1}{G-1},
\frac{2}{G-1}
\right]
}
\label{eq:practical_gaussian_interval}
\end{equation}
as the common reference range for comparing GKAN and PU-GKAN. This interval is a practical conditioning-based range rather than a claim that either endpoint is universally optimal; further derivation and numerical evidence are given in~\cite{Amir_GKAN}.

\section{Numerical Examples}
\label{Examples}

\subsection{Target Functions}

\begin{figure}[hbt!]
\centering
\includegraphics[width=6.5in]{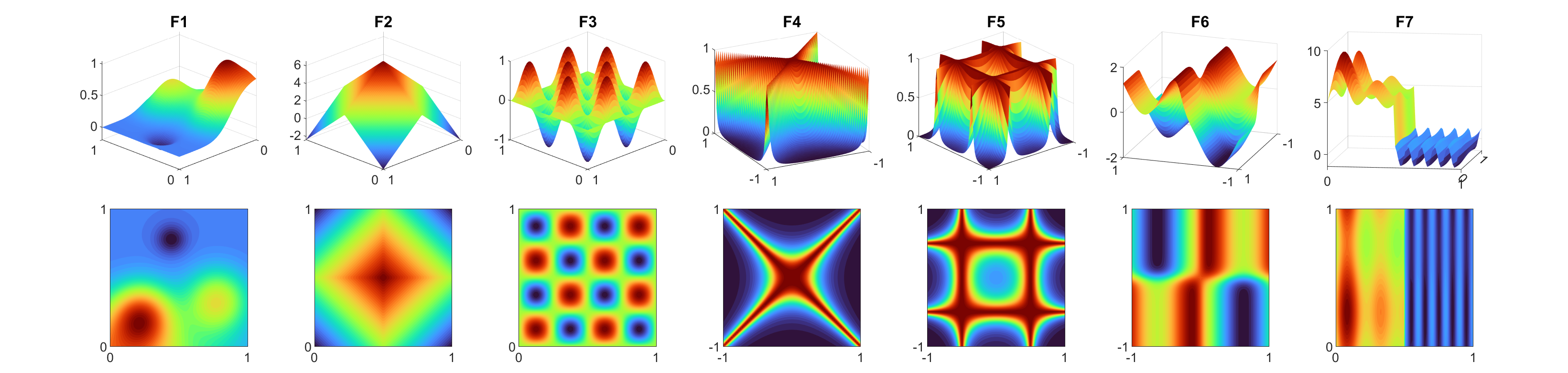}
\caption{Target functions used in this study for validation.}
\label{target_functions3}
\end{figure}

We consider seven two-dimensional target functions defined on either \([0,1]^2\) or \([-1,1]^2\). 
They are denoted by \(F1,\ldots,F7\), plotted in Fig.~\ref{target_functions3}, and listed below.

\begin{align}\label{F1}
F1(x,y)
&=
0.75\exp\!\left(-\frac{(9x-2)^2+(9y-2)^2}{4}\right)
+
0.75\exp\!\left(-\frac{(9x+1)^2}{49}-\frac{(9y+1)^2}{10}\right)
\nonumber\\
&\quad
+
0.5\exp\!\left(-\frac{(9x-7)^2+(9y-3)^2}{4}\right)
-
0.2\exp\!\left(-(9x-4)^2-(9y-7)^2\right),
\qquad (x,y)\in[0,1]^2,
\\[0.5em]
F2(x,y)
&=
\frac{1}{9}\left(64-81\left(|x-\tfrac12|+|y-\tfrac12|\right)\right)-\frac12,
\qquad (x,y)\in[0,1]^2,
\\[0.5em]
F3(x,y)
&=
\sin(4\pi x)\sin(4\pi y),
\qquad (x,y)\in[0,1]^2,
\\[0.5em]
F4(x,y)
&=
\frac{1}{1+100(x^2-y^2)^2},
\qquad (x,y)\in[-1,1]^2,
\\[0.5em]
F5(x,y)
&=
\frac{1}{1+10^3\left((x^2-0.25)^2(y^2-0.25)^2\right)},
\qquad (x,y)\in[-1,1]^2,
\\[0.5em]
F6(x,y)
&=
\frac{\tanh(10x)\tanh(10y)}{\tanh^2(10)}+\cos(5x),
\qquad (x,y)\in[-1,1]^2.
\end{align}

For the discontinuous target, define
\begin{equation}
s(x)
=
\sin(2\pi x)+\sin(4\pi x)+\sin(6\pi x)+\sin(8\pi x).
\end{equation}
Then
\begin{equation}
F7(x,y)
=
\begin{cases}
\left(5+s(x)\right)\left(1+0.15\sin(2\pi y)\right),
& x<\frac12, \\[0.4em]
\cos(20\pi x)\left(1+0.15\sin(2\pi y)\right),
& x\ge \frac12,
\end{cases}
\qquad (x,y)\in[0,1]^2.
\end{equation}

\subsection{Experimental Setup and Error Metric}

All input variables are affinely normalized to \([0,1]\) before being passed to the first KAN layer. 
The Gaussian centers are fixed on this normalized interval, and errors are evaluated in the original physical coordinates. 
Unless stated otherwise, only the trainable edge coefficients are optimized; the Gaussian scale parameter \(\epsilon\) is fixed during each training run. 
All models are trained using the AdamW optimizer with learning rate \(10^{-3}\). 

For the scale-sensitivity experiments, the Gaussian scale parameter is swept over
\[
\epsilon \in [0.005,10],
\]
using \(100\) logarithmically spaced values. 
For each value of \(\epsilon\), the model is trained for \(10{,}000\) epochs, and the validation error is recorded. 
Training points are generated using Halton sequences, which are commonly used in meshfree and quasi-Monte Carlo approximation settings~\cite{Fasshauer07}. 
For the two-dimensional function-approximation examples, validation is performed on a uniform \(90\times 90\) grid. 
In these experiments, \(N\) denotes the number of collocation points used for function approximation problems, while for PDE problems it denotes the total number of training points, including interior, boundary, and initial-condition points when applicable. 
The parameter \(G\) denotes the number of Gaussian grid centers used in each edge basis. 
Unless stated otherwise, the results reported in the figures and tables are geometric averages over at least four random seeds.

The validation error is measured by the root-mean-square error (RMSE),
\begin{equation}
\mathrm{RMSE}
=
\left(
\frac{1}{N_{\mathrm{val}}}
\sum_{i=1}^{N_{\mathrm{val}}}
\left|
u_{\theta}(\mathbf{x}_i)-u(\mathbf{x}_i)
\right|^2
\right)^{1/2},
\label{eq:rmse}
\end{equation}
where \(u_{\theta}\) is the model prediction, \(u\) is the exact target function or exact PDE solution, and \(N_{\mathrm{val}}\) is the number of validation points.

For each experiment, the optimal Gaussian scale is selected as the value that gives the smallest validation RMSE. 
For the standard Gaussian KAN, we denote this value by
\begin{equation}
\epsilon_{GKAN}^{*}
=
\arg\min_{\epsilon}
\mathrm{RMSE}_{GKAN}(\epsilon),
\label{eq:eps_star_gaussian}
\end{equation}
while for the Shepard-normalized Gaussian KAN, we denote it by
\begin{equation}
\epsilon_{PU-GKAN}^{*}
=
\arg\min_{\epsilon}
\mathrm{RMSE}_{PU-GKAN}(\epsilon).
\label{eq:eps_star_normalized_gaussian}
\end{equation}
Here, \(\mathrm{RMSE}_{GKAN}(\epsilon)\) and \(\mathrm{RMSE}_{PU-GKAN}(\epsilon)\) are the validation RMSE values obtained by the standard Gaussian and Shepard-normalized Gaussian models, respectively, at scale \(\epsilon\).

The percentage improvement reported in the tables is computed as
\begin{equation}
\mathrm{Imp.}
=
\frac{
\mathrm{RMSE}_{GKAN}
-
\mathrm{RMSE}_{PU-GKAN}
}{
\mathrm{RMSE}_{GKAN}
}
\times 100\%.
\label{eq:improvement}
\end{equation}
Thus, positive values indicate that the Shepard-normalized model is more accurate, while negative values indicate that the standard Gaussian KAN gives the lower RMSE.

\subsection{Representative Error Curves and Computational Time}
\label{subsec:representative_error_curves}

\begin{figure}[!ht]
\centering
\includegraphics[width=6.0in]{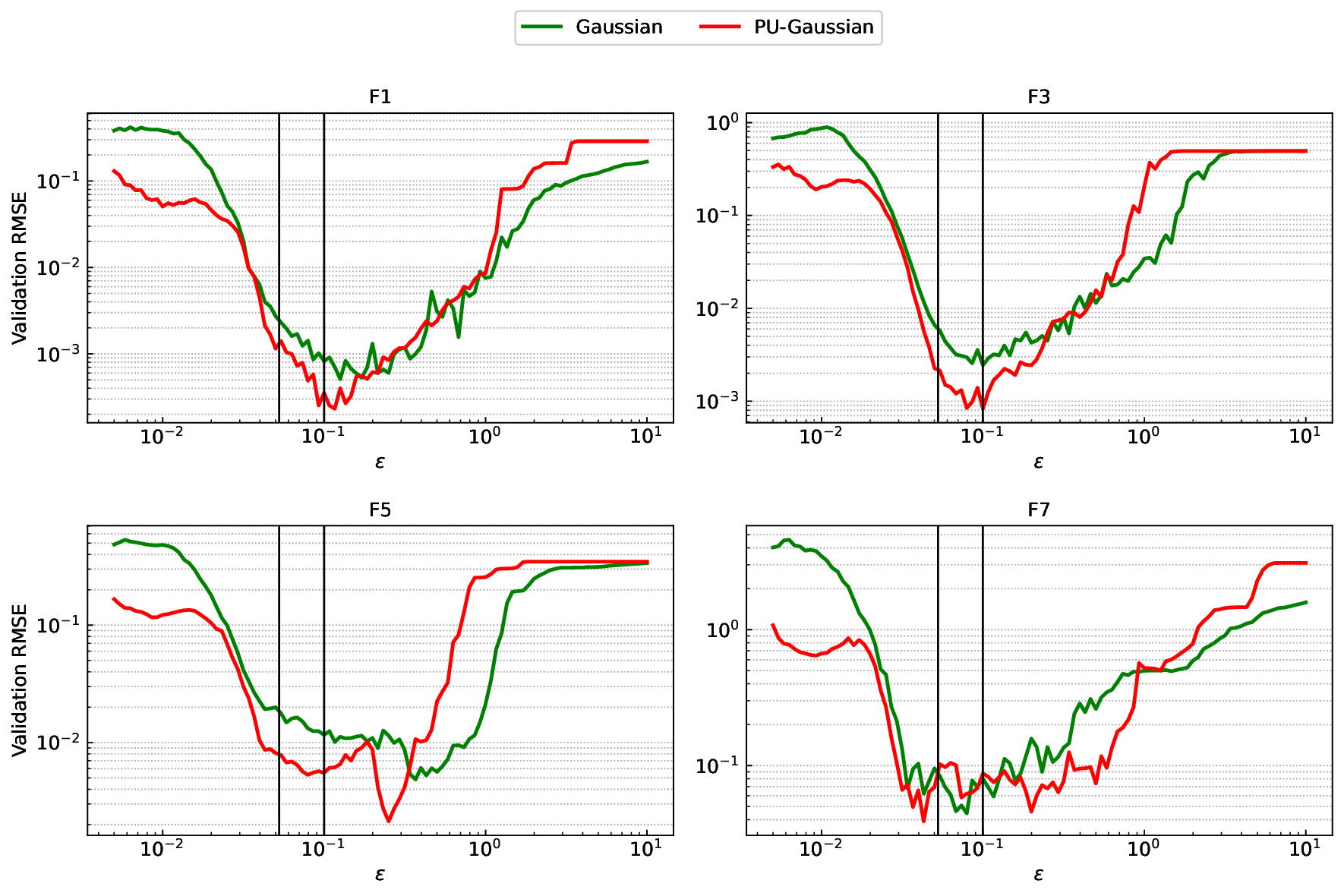}
\caption{Validation RMSE versus \(\epsilon\) for \(F1\), \(F3\), \(F5\), and \(F7\), comparing the Gaussian KAN and the Shepard-normalized Gaussian KAN with \(N=1000\) and \(G=20\). The vertical lines mark \(\epsilon=1/(G-1)\) and \(\epsilon=2/(G-1)\).}
\label{fig:representative_eps_sweeps}
\end{figure}

We first compare the standard Gaussian KAN and the Shepard-normalized Gaussian KAN on four representative targets before presenting the full parameter sweeps. 
Figure~\ref{fig:representative_eps_sweeps} reports the validation RMSE as a function of \(\epsilon\) for \(F1\), \(F3\), \(F5\), and \(F7\), with \(N=1000\) and \(G=20\). 
The vertical lines mark the reference scales \(\epsilon=1/(G-1)\) and \(\epsilon=2/(G-1)\).

Before discussing the accuracy trends, we also note that the additional computational cost of the Shepard-normalized model is modest. 
Both models evaluate the same Gaussian basis functions and use the same coefficient contraction; the only extra operations in the Shepard-normalized version are the summation of the Gaussian basis values over the grid points and the subsequent normalization. 
Thus, the two models have the same asymptotic computational structure, while the normalized model has a slightly larger constant cost. 
In a representative Google Colab GPU benchmark, PU-GKAN required about \(1.25\times\) the forward time and \(1.16\times\) the training-step time of the standard GKAN. 
In the full training runs, where validation and other loop overheads are also included, the observed total-time ratio remained modest and was approximately \(1.19\times\) on average over four representative seeds.

The error curves show the expected sensitivity to \(\epsilon\): very small and very large scales are generally unfavorable. 
For \(F1\), \(F3\), and \(F5\), Shepard normalization gives lower error over a broad part of the relevant range and usually a lower minimum. 
For the discontinuous target \(F7\), the improvement is less uniform, which is expected since discontinuities are harder to represent with smooth radial bases. 
These examples motivate the systematic comparisons with respect to \(N\), \(G\), architecture, and PDE settings in the following subsections.

The computational overhead was evaluated over \(5000\) epochs using float32 on an NVIDIA T4 GPU provided by Google Colab in 2026. The four settings are
\[
\begin{aligned}
\mathcal{S}_1&=(G,N,\mathcal{A})=(20,2000,(2,12,12,1)),&
\mathcal{S}_2&=(10,2000,(2,12,12,1)),\\
\mathcal{S}_3&=(20,4000,(2,12,12,1)),&
\mathcal{S}_4&=(20,4000,(2,12,12,12,12,1)).
\end{aligned}
\]
Table~\ref{tab:computational_cost} reports the increase in average time per epoch relative to GKAN. The PU-GKAN overhead ranges from \(20.1\%\) to \(24.1\%\). For example, under \(\mathcal{S}_1\), the total time for \(5000\) epochs increases from \(19.48\)~s for GKAN to \(23.84\)~s for PU-GKAN.

\begin{table}[hbt!]
\centering
\caption{Increase in average time per epoch relative to GKAN over \(5000\) epochs on an NVIDIA T4 GPU.}
\label{tab:computational_cost}
\begin{tabular}{lcccc}
\hline
Method
& \(\mathcal{S}_1\)
& \(\mathcal{S}_2\)
& \(\mathcal{S}_3\)
& \(\mathcal{S}_4\)\\
\hline
GKAN
& Reference
& Reference
& Reference
& Reference\\
PU-GKAN
& \(22.4\%\)
& \(22.9\%\)
& \(20.1\%\)
& \(24.1\%\)\\
\hline
\end{tabular}
\end{table}

\subsection{Effect of the Sample Size \(N\)}
\label{subsec:effect_sample_size}

We next study the effect of the number of training samples \(N\) at fixed \(G=20\). 
For each target and each value of \(N\), Table~\ref{tab:best_rmse_gaussian_vs_normalized_N} reports the best validation RMSE over the reference interval
\[
\epsilon\in\left[\frac{1}{G-1},\frac{2}{G-1}\right].
\]
The reported errors are obtained by first taking the geometric mean over seeds at each \(\epsilon\), and then selecting the best \(\epsilon\) in this interval.

The Shepard-normalized Gaussian KAN gives lower RMSE for most targets and sample sizes. 
The improvement is especially clear for the smooth and moderately regular targets \(F1\), \(F2\), \(F3\), \(F4\), \(F5\), and \(F6\). 
The discontinuous target \(F7\) is less uniform: the standard Gaussian KAN is better for \(N=500\) and \(N=1500\), while the normalized model becomes better at \(N=2000\).

\begin{figure}[!ht]
\centering
\includegraphics[width=6.5in]{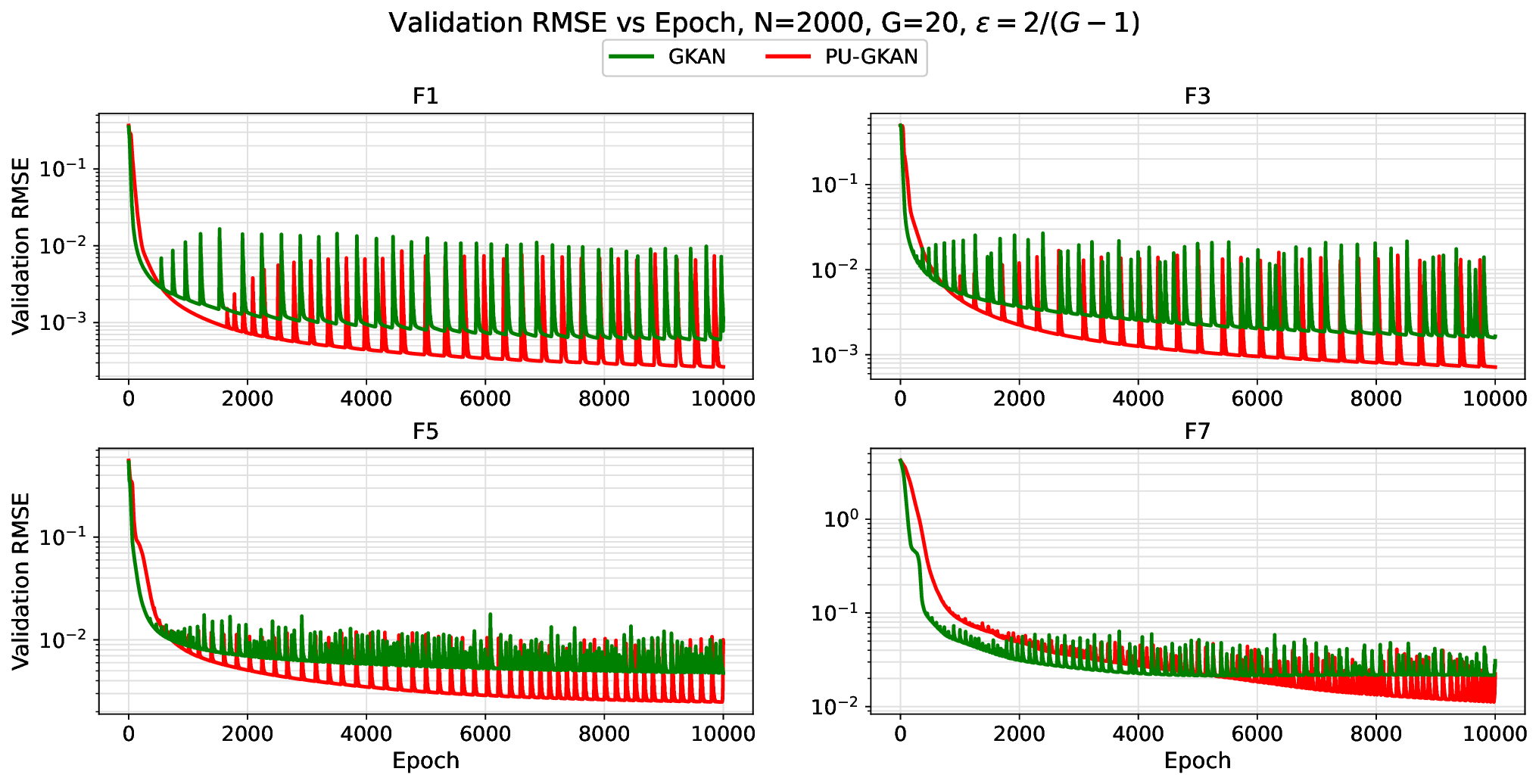}
\caption{Validation RMSE versus epoch for representative targets with \(N=2000\), \(G=20\), and \(\epsilon=0.1\).}
\label{fig:val_rmse_epoch_N_sweep}
\end{figure}

Figure~\ref{fig:val_rmse_epoch_N_sweep} shows the validation RMSE during training for representative targets at \(N=2000\), \(G=20\), and \(\epsilon=0.1\). 
The normalized model generally converges to a lower error and remains below the standard Gaussian KAN for most of the training process, although the improvement is less pronounced for the discontinuous target.

\begin{table*}[t]
\centering
\caption{Best validation RMSE over \(\epsilon\in[1/(G-1),2/(G-1)]\) for the Gaussian KAN (GKAN) and the partition-of-unity Gaussian KAN (PU-GKAN) at fixed \(G=20\). Positive improvement means that PU-GKAN is more accurate.}
\label{tab:best_rmse_gaussian_vs_normalized_N}
\begin{tabular}{cc c c c c c}
\toprule
F & $N$ & $\epsilon_{\mathrm{GKAN}}^*$ & $\epsilon_{\mathrm{PU-GKAN}}^*$ & GKAN & PU-GKAN & Imp. \\
\midrule
\multirow{3}{*}{F1} & 500 & 0.100 & 0.092 & 1.95e-03 & \textbf{5.05e-04} & 74.1\% \\
 & 1500 & 0.073 & 0.100 & 6.22e-04 & \textbf{2.68e-04} & 56.9\% \\
 & 2000 & 0.068 & 0.068 & 6.11e-04 & \textbf{3.71e-04} & 39.3\% \\
\midrule
\multirow{3}{*}{F2} & 500 & 0.100 & 0.079 & 4.00e-02 & \textbf{1.73e-02} & 56.8\% \\
 & 1500 & 0.058 & 0.073 & 1.27e-02 & \textbf{6.27e-03} & 50.7\% \\
 & 2000 & 0.054 & 0.063 & 1.02e-02 & \textbf{5.02e-03} & 50.8\% \\
\midrule
\multirow{3}{*}{F3} & 500 & 0.100 & 0.092 & 8.24e-03 & \textbf{2.05e-03} & 75.1\% \\
 & 1500 & 0.092 & 0.086 & 2.13e-03 & \textbf{7.79e-04} & 63.5\% \\
 & 2000 & 0.092 & 0.092 & 1.53e-03 & \textbf{7.32e-04} & 52.3\% \\
\midrule
\multirow{3}{*}{F4} & 500 & 0.068 & 0.092 & 6.81e-02 & \textbf{4.35e-02} & 36.2\% \\
 & 1500 & 0.100 & 0.079 & 1.85e-02 & \textbf{6.11e-03} & 67.0\% \\
 & 2000 & 0.100 & 0.079 & 1.05e-02 & \textbf{4.17e-03} & 60.3\% \\
\midrule
\multirow{3}{*}{F5} & 500 & 0.100 & 0.100 & 2.55e-02 & \textbf{1.44e-02} & 43.7\% \\
 & 1500 & 0.086 & 0.079 & 4.99e-03 & \textbf{2.84e-03} & 43.1\% \\
 & 2000 & 0.086 & 0.054 & 3.88e-03 & \textbf{1.83e-03} & 52.9\% \\
\midrule
\multirow{3}{*}{F6} & 500 & 0.100 & 0.086 & 7.71e-03 & \textbf{1.68e-03} & 78.2\% \\
 & 1500 & 0.100 & 0.092 & 1.83e-03 & \textbf{8.53e-04} & 53.5\% \\
 & 2000 & 0.092 & 0.073 & 1.49e-03 & \textbf{7.22e-04} & 51.4\% \\
\midrule
\multirow{3}{*}{F7} & 500 & 0.086 & 0.073 & \textbf{7.47e-02} & 9.17e-02 & -22.7\% \\
 & 1500 & 0.073 & 0.073 & \textbf{1.22e-02} & 1.84e-02 & -50.8\% \\
 & 2000 & 0.079 & 0.054 & 1.02e-02 & \textbf{7.02e-03} & 31.2\% \\
\bottomrule
\end{tabular}%
\end{table*}

\subsection{Effect of the Number of Centers \(G\)}
\label{subsec:effect_num_centers}

We next vary the number of Gaussian centers \(G\) at fixed \(N=1000\). 
For each target and each value of \(G\), Table~\ref{tab:best_rmse_gaussian_vs_normalized_G} reports the best validation RMSE over the corresponding range
\[
\epsilon\in\left[\frac{1}{G-1},\frac{2}{G-1}\right],
\]
where the range changes with \(G\). 
As before, errors are obtained by taking the geometric mean across seeds at each \(\epsilon\), followed by selecting the best \(\epsilon\) in the admissible interval.

The normalized Gaussian KAN gives lower RMSE for all tested values of \(G\) on \(F1\), \(F2\), \(F3\), \(F4\), \(F5\), and \(F6\). 
The gains remain clear as \(G\) increases, showing that the benefit of normalization is not limited to small feature dictionaries. 
For the discontinuous target \(F7\), the standard Gaussian KAN remains better for all tested \(G\), indicating that the normalized basis is less effective for this case.

\begin{table*}[t]
\centering
\caption{Best validation RMSE over \(\epsilon\in[1/(G-1),2/(G-1)]\) for the Gaussian KAN (GKAN) and the partition-of-unity Gaussian KAN (PU-GKAN) at fixed \(N=1000\). The admissible \(\epsilon\)-range is recomputed for each \(G\). Positive improvement means that PU-GKAN is more accurate.}
\label{tab:best_rmse_gaussian_vs_normalized_G}
\begin{tabular}{cc cc ccc}
\toprule
F & $G$ & $\epsilon_{\mathrm{GKAN}}^*$ & $\epsilon_{\mathrm{PU-GKAN}}^*$ & GKAN & PU-GKAN & Imp. \\
\midrule
\multirow{3}{*}{F1} & 14 & 0.126 & 0.116 & 4.37e-04 & \textbf{2.51e-04} & 42.6\% \\
 & 18 & 0.108 & 0.108 & 4.78e-04 & \textbf{2.10e-04} & 56.1\% \\
 & 22 & 0.079 & 0.092 & 5.55e-04 & \textbf{2.16e-04} & 61.2\% \\
\midrule
\multirow{3}{*}{F2} & 14 & 0.116 & 0.116 & 7.30e-03 & \textbf{4.95e-03} & 32.1\% \\
 & 18 & 0.100 & 0.068 & 7.00e-03 & \textbf{5.62e-03} & 19.7\% \\
 & 22 & 0.079 & 0.063 & 8.14e-03 & \textbf{5.88e-03} & 27.7\% \\
\midrule
\multirow{3}{*}{F3} & 14 & 0.100 & 0.108 & 1.88e-03 & \textbf{9.97e-04} & 47.1\% \\
 & 18 & 0.092 & 0.100 & 1.88e-03 & \textbf{8.77e-04} & 53.3\% \\
 & 22 & 0.079 & 0.086 & 1.90e-03 & \textbf{6.38e-04} & 66.5\% \\
\midrule
\multirow{3}{*}{F4} & 14 & 0.147 & 0.108 & 1.17e-02 & \textbf{4.92e-03} & 58.0\% \\
 & 18 & 0.100 & 0.100 & 1.38e-02 & \textbf{4.98e-03} & 63.9\% \\
 & 22 & 0.092 & 0.079 & 2.66e-02 & \textbf{7.25e-03} & 72.7\% \\
\midrule
\multirow{3}{*}{F5} & 14 & 0.100 & 0.086 & 6.34e-03 & \textbf{4.62e-03} & 27.1\% \\
 & 18 & 0.073 & 0.068 & 7.69e-03 & \textbf{3.98e-03} & 48.2\% \\
 & 22 & 0.073 & 0.050 & 7.67e-03 & \textbf{2.92e-03} & 62.0\% \\
\midrule
\multirow{3}{*}{F6} & 14 & 0.136 & 0.108 & 2.01e-03 & \textbf{9.89e-04} & 50.9\% \\
 & 18 & 0.086 & 0.092 & 1.89e-03 & \textbf{7.81e-04} & 58.7\% \\
 & 22 & 0.092 & 0.092 & 2.15e-03 & \textbf{7.70e-04} & 64.2\% \\
\midrule
\multirow{3}{*}{F7} & 14 & 0.108 & 0.086 & \textbf{6.20e-02} & 6.64e-02 & -7.1\% \\
 & 18 & 0.086 & 0.073 & \textbf{4.58e-02} & 8.52e-02 & -86.0\% \\
 & 22 & 0.086 & 0.050 & \textbf{4.92e-02} & 6.34e-02 & -28.8\% \\
\bottomrule
\end{tabular}%
\end{table*}

\subsection{Extension to Matérn Bases}
\label{subsec:matern_basis_results}

We also test whether the same normalization idea applies beyond Gaussian bases. 
Here we use the Matérn-5 basis
\[
\phi_{\mathrm{M5}}(\rho;\epsilon)
=
\left(1+r+\frac{r^2}{3}\right)e^{-r},
\qquad
r=\frac{\sqrt{10}\,\rho}{\epsilon},
\]
where \(\rho\) denotes the distance to a center. 
The lower reference scale is chosen by the same adjacent-center overlap rule used for the Gaussian case. 
With \(h=1/(G-1)\) and overlap level \(e^{-1}\), we solve
\[
\left(1+s+\frac{s^2}{3}\right)e^{-s}=e^{-1},
\qquad s\approx 2.90463,
\]
which gives
\[
\epsilon_{\mathrm{low}}^{\mathrm{M5}}
=
\frac{\sqrt{10}}{2.90463}h
\approx
\frac{1.09}{G-1}.
\]
The upper reference scale is determined by the first-layer conditioning boundary \(\kappa(A_\epsilon)\approx 3\times 10^3\). 
In the experiments, this boundary is typically farther to the right than in the Gaussian case, indicating that the Matérn basis remains better conditioned for larger \(\epsilon\).

\begin{figure}[!ht]
\centering
\includegraphics[width=6.0in]{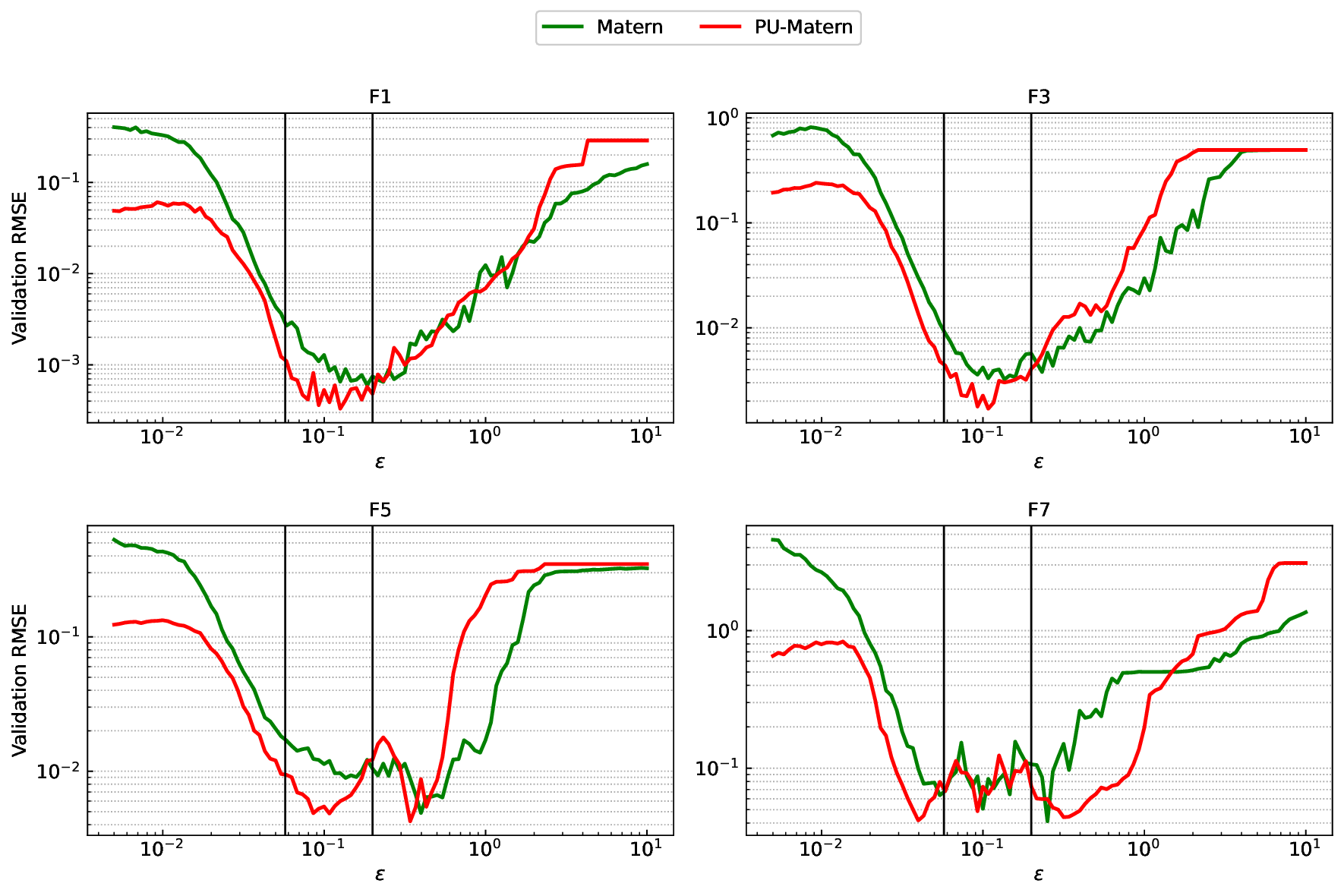}
\caption{Validation RMSE versus \(\epsilon\) for the Matérn KAN and Shepard-normalized Matérn KAN with \(N=1000\) and \(G=20\). The vertical lines mark \(\epsilon_{\mathrm{low}}^{\mathrm{M5}}\approx 1.09/(G-1)\) and the first-layer conditioning boundary \(\kappa(A_\epsilon)\approx 3\times10^3\).}
\label{fig:matern_eps_sweeps}
\end{figure}

Figure~\ref{fig:matern_eps_sweeps} compares the Matérn KAN and its Shepard-normalized variant for \(F1\), \(F3\), \(F5\), and \(F7\). 
The normalized Matérn model generally lowers the error for the smooth and moderately regular targets, while the discontinuous case remains less uniform. 
This confirms that the benefit of Shepard-type normalization is not restricted to Gaussian bases.

\newpage
\pagebreak
\subsection{Higher Dimensions and Different Architectures}
\label{subsec:high_dimension_architecture}

We next examine the dimension-dependent test problem
\begin{equation}
f_d(\mathbf{x})
=
\exp\!\left(
\frac{1}{d}\sum_{i=1}^{d}
\left(
\sin(\pi x_i)+\frac{1}{2}x_i^2
\right)
\right),
\qquad
\mathbf{x}\in[0,1]^d .
\end{equation}
Training and evaluation points are generated on \([0,1]^d\) using Halton sequences. 
We compare the Gaussian KAN with its Shepard-normalized variant over four representative settings, where the dimension \(d\), sample size \(N\), and architecture are varied together.

\begin{figure}[!ht]
\centering
\includegraphics[width=6.0in]{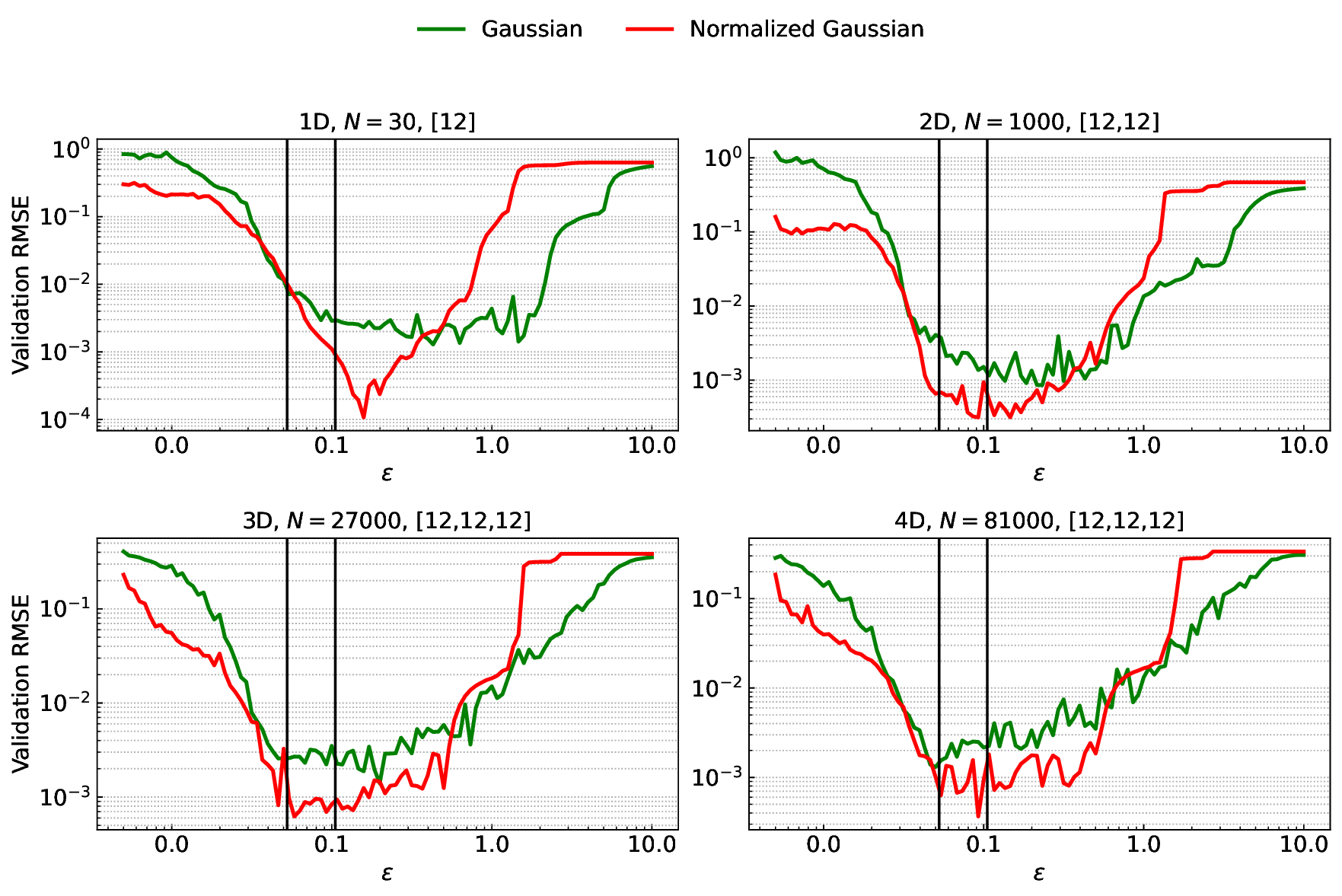}
\caption{Validation RMSE versus \(\epsilon\) for the Gaussian and Shepard-normalized Gaussian KANs across four dimension--architecture settings. The vertical reference lines mark \(\epsilon=1/(G-1)\) and \(\epsilon=2/(G-1)\).}
\label{fig:hd_val_rmse_vs_eps}
\end{figure}

Figure~\ref{fig:hd_val_rmse_vs_eps} shows the validation RMSE as a function of \(\epsilon\). 
In all cases, the error curves are U-shaped, so both overly small and overly large values of \(\epsilon\) are unfavorable. 
The two vertical reference lines indicate the interval
\[
\left[\frac{1}{G-1},\frac{2}{G-1}\right].
\]
Across all four settings, the normalized Gaussian KAN attains a lower minimum RMSE and usually maintains a broader low-error region.

Table~\ref{tab:hd_best_eps_gaussian_vs_normalized} reports the best validation RMSE in this interval. 
The normalized model is more accurate in every tested case, with improvements ranging from \(61.5\%\) to \(77.0\%\). 
These results show that the benefit of Shepard-type normalization is not restricted to the low-dimensional examples considered earlier and remains clear across the tested dimension--architecture pairs.

\begin{table}[t]
\centering
\caption{Best validation RMSE over \(\epsilon\in[1/(G-1),\,2/(G-1)]\) for the Gaussian KAN (GKAN) and the partition-of-unity Gaussian KAN (PU-GKAN) in the tested dimension--architecture settings. Positive improvement means that PU-GKAN is more accurate.}
\label{tab:hd_best_eps_gaussian_vs_normalized}
\begin{tabular}{cc c cc ccc}
\toprule
$d$ & $N$ & Arch. & $\epsilon_{\mathrm{GKAN}}^*$ & $\epsilon_{\mathrm{PU-GKAN}}^*$ & GKAN & PU-GKAN & Imp. \\
\midrule
1 & 30 & [12] & 0.100 & 0.100 & 2.88e-03 & \textbf{1.11e-03} & 61.5\% \\
2 & 1000 & [12,12] & 0.092 & 0.092 & 1.37e-03 & \textbf{3.15e-04} & 77.0\% \\
3 & 27000 & [12,12,12] & 0.092 & 0.058 & 2.22e-03 & \textbf{6.22e-04} & 72.0\% \\
4 & 81000 & [12,12,12] & 0.054 & 0.092 & 1.55e-03 & \textbf{3.66e-04} & 76.4\% \\
\bottomrule
\end{tabular}
\end{table}

\subsection{Physics-Informed Helmholtz Problem}
\label{subsec:helmholtz_equation}

We next consider the two-dimensional Helmholtz problem
\begin{equation}
-\Delta u-\lambda u=f
\qquad \text{in } \Omega=(0,1)^2,
\label{eq:pde_model}
\end{equation}
with homogeneous Dirichlet boundary condition
\begin{equation}
u=0
\qquad \text{on } \partial\Omega,
\label{eq:pde_bc}
\end{equation}
and fix
\[
\lambda=100.
\]
As a reference solution, we take
\begin{equation}
u(x,y)=\sin(a_1\pi x)\sin(a_2\pi y),
\qquad (x,y)\in\Omega,
\label{eq:pde_exact}
\end{equation}
with \((a_1,a_2)=(1,4)\). 
Substituting \eqref{eq:pde_exact} into \eqref{eq:pde_model} gives
\begin{equation}
f(x,y)
=
\bigl((a_1^2+a_2^2)\pi^2-\lambda\bigr)\,u(x,y).
\label{eq:pde_forcing}
\end{equation}

Both models use the architecture \((2,12,12,1)\) with \(G=20\) Gaussian centers per layer. 
We sample \(N=2000\) interior Halton points and \(200\) boundary points, and evaluate the solution on a \(90\times 90\) grid. 
The boundary-condition weight is \(W_{\mathrm{BC}}=100\). 
We compare the standard Gaussian KAN and the Shepard-normalized Gaussian KAN by sweeping \(\epsilon\) over \([0.005,10]\).

\begin{figure}[hbt!]
\centering%
\subfigure[]{
\label{rmse_vs_eps_L100_N2000_G20_A14_W100_T90}
\includegraphics[width=3in]{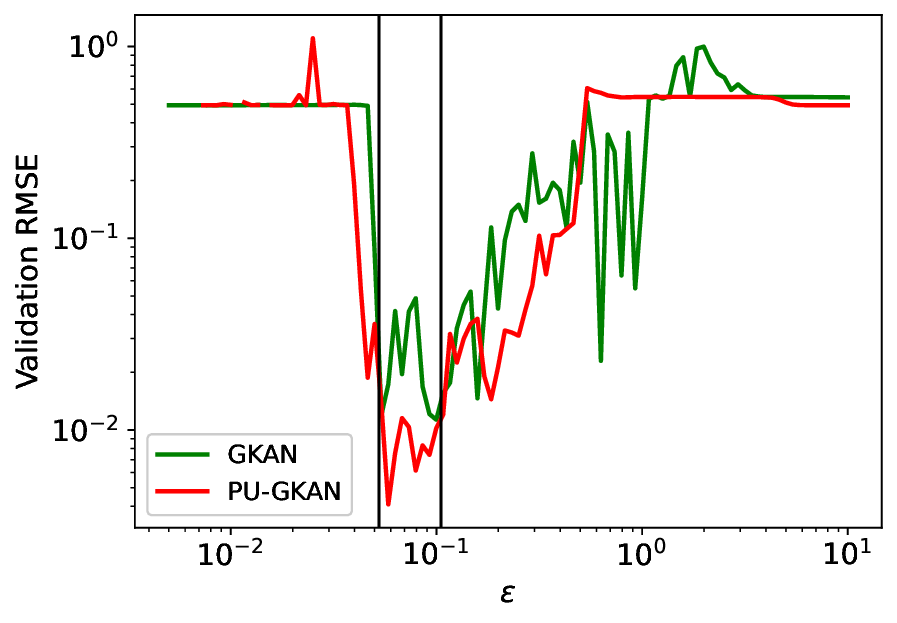}}
\subfigure[]{
\label{rmse_vs_epoch_L100_N2000_G20_A14_W100_T90_S0}
\includegraphics[width=3in]{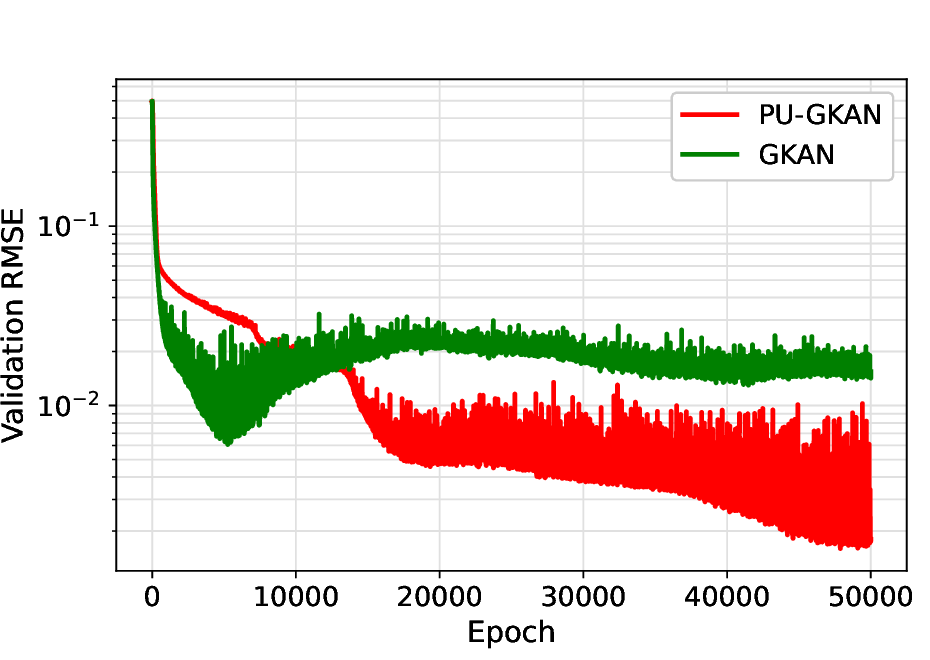}}
\caption{Helmholtz problem with \(\lambda=100\): (left) validation RMSE versus \(\epsilon\); (right) validation RMSE versus epoch for a representative run at \(\epsilon=2/(G-1)\).}
\label{Ex1_domain}
\end{figure}

Figure~\ref{Ex1_domain} summarizes the results. 
In Fig.~\ref{rmse_vs_eps_L100_N2000_G20_A14_W100_T90}, the normalized model attains a lower minimum validation RMSE and remains favorable over the useful range of \(\epsilon\). 
The advantage is most visible near the optimal region, where the standard Gaussian model reaches its best performance only over a narrower interval. 
Figure~\ref{rmse_vs_epoch_L100_N2000_G20_A14_W100_T90_S0} shows the validation RMSE during training for one representative run at \(\epsilon=2/(G-1)\). 
After the initial stage, the normalized model continues to decrease and stays clearly below the standard Gaussian model.

These results show that the benefit of Shepard-type normalization persists in this more challenging PDE setting. 
For \(\lambda=100\), normalization improves both the \(\epsilon\)-sensitivity and the final accuracy.

\begin{figure}[hbt!]
\centering
\includegraphics[width=5.5in]{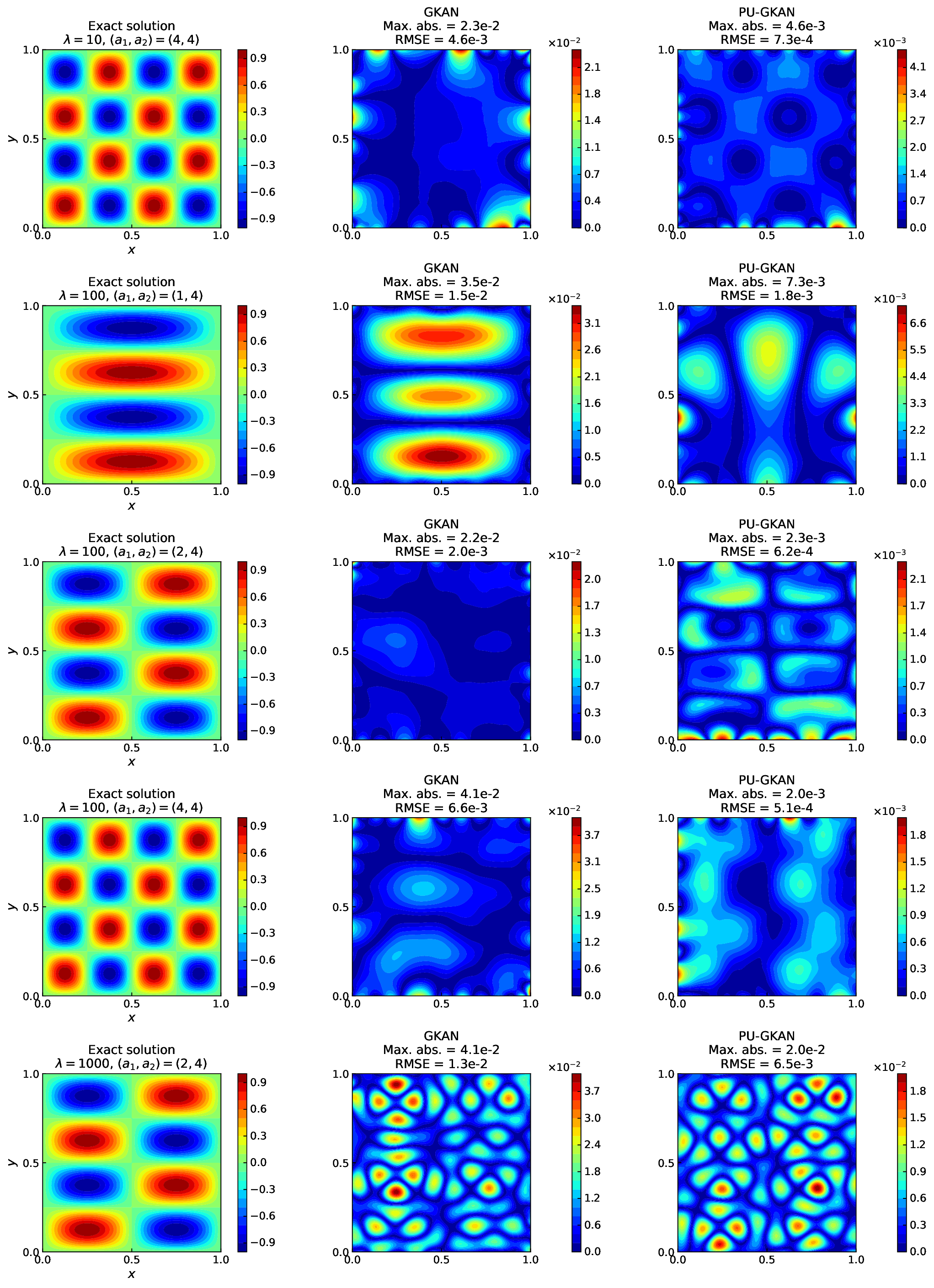}
\caption{Exact solutions and absolute-error fields at the final epoch of \(50{,}000\) training epochs with \(\epsilon=2/(G-1)\). From top to bottom, \((\lambda,a_1,a_2)=(10,4,4)\), \((100,1,4)\), \((100,2,4)\), \((100,4,4)\), and \((1000,2,4)\). The columns show the exact solution, the GKAN absolute error, and the PU-GKAN absolute error. The maximum absolute error and validation RMSE are reported above each error field.}
\label{all_l}
\end{figure}

We extend the comparison to
\[
(\lambda,a_1,a_2)
=
(10,4,4),\,
(100,1,4),\,
(100,2,4),\,
(100,4,4),\,
(1000,2,4).
\]
All numerical settings remain unchanged: both models use the architecture \((2,12,12,1)\), \(G=20\) Gaussian centers per layer, \(2000\) interior Halton collocation points, \(200\) boundary points, a \(90\times90\) evaluation grid, and \(W_{\mathrm{BC}}=100\). The shape parameter is fixed at \(\epsilon=2/(G-1)\). Each model is trained for \(50{,}000\) epochs, and Fig.~\ref{all_l} presents the results at the final epoch.

PU-GKAN gives a lower RMSE and maximum absolute error in all five cases. Relative to GKAN, the minimum and maximum reductions in RMSE are \(50.0\%\) and \(92.3\%\), respectively, while the minimum and maximum reductions in the maximum absolute error are \(51.2\%\) and \(95.1\%\), respectively. Thus, PU-GKAN improves the final-epoch accuracy for every parameter combination considered.

\subsection{Physics-Informed Wave Equation}
\label{sec:wave_equation}

We next consider the one-dimensional wave equation on the space--time domain
\[
(x,t)\in [0,1]\times[0,3].
\]
The governing equation is
\begin{equation}
u_{tt}(x,t)-u_{xx}(x,t)=0,
\qquad (x,t)\in (0,1)\times(0,3),
\label{eq:wave_pde}
\end{equation}
with homogeneous Dirichlet boundary conditions
\begin{equation}
u(0,t)=u(1,t)=0,
\qquad t\in[0,3],
\label{eq:wave_bc}
\end{equation}
and initial conditions
\begin{equation}
u(x,0)=\frac{1}{2}\sin(\pi x),
\label{eq:wave_ic_u}
\end{equation}
\begin{equation}
u_t(x,0)=\pi \sin(3\pi x).
\label{eq:wave_ic_ut}
\end{equation}
These data are induced by the exact solution
\begin{equation}
u(x,t)
=
\frac{1}{2}\sin(\pi x)\cos(\pi t)
+
\frac{1}{3}\sin(3\pi x)\sin(3\pi t),
\qquad (x,t)\in[0,1]\times[0,3],
\label{eq:wave_exact}
\end{equation}
which satisfies \eqref{eq:wave_pde}--\eqref{eq:wave_ic_ut}.

\begin{figure}[hbt!]
\centering
\subfigure[]{
\label{fig:wave_val_rmse_vs_eps}
\includegraphics[width=3in]{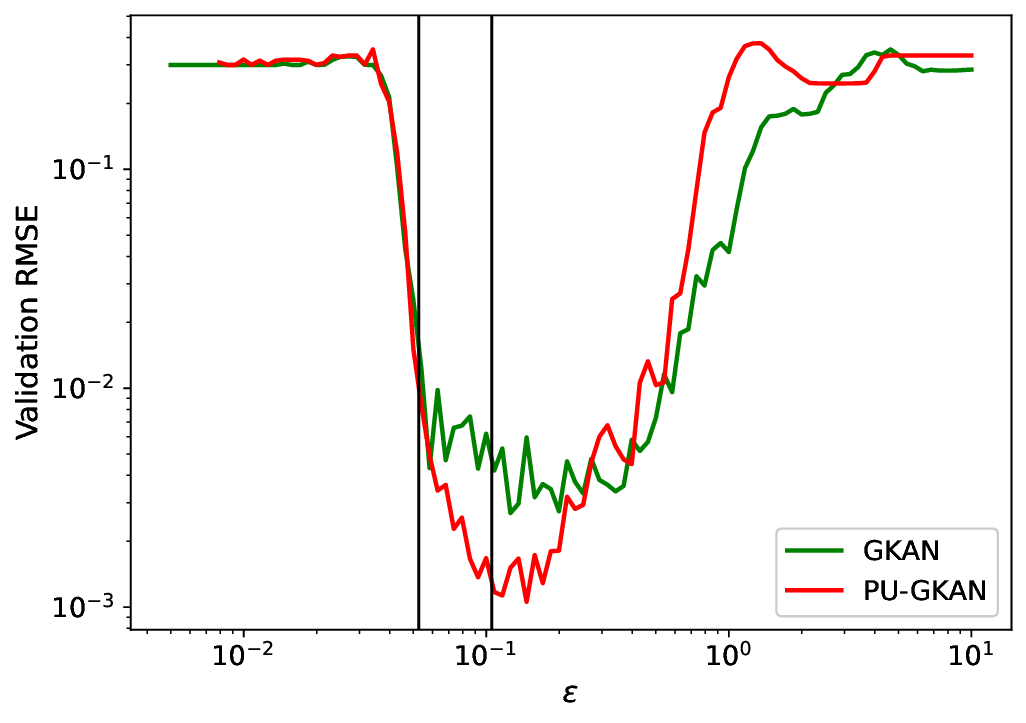}}
\subfigure[]{
\label{fig:wave_val_rmse_vs_epoch}
\includegraphics[width=2.8in]{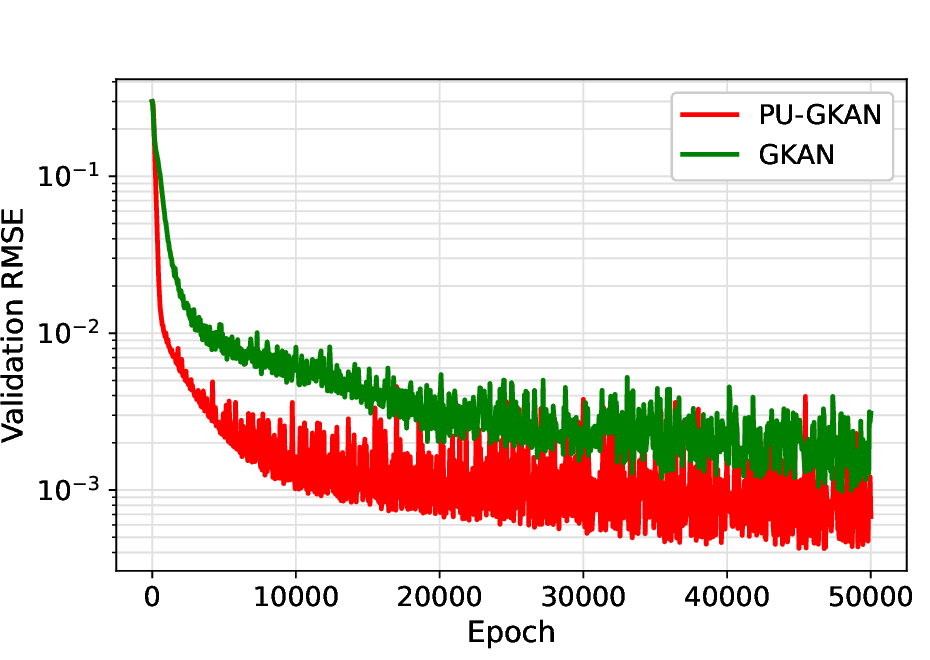}}
\caption{Wave-equation results for the Gaussian KAN and the Shepard-normalized Gaussian KAN. (a) Validation RMSE versus \(\epsilon\) after \(10{,}000\) training epochs for each value of \(\epsilon\), with \(N=6000\) and \(G=20\). (b) Validation RMSE versus training epoch at \(\epsilon=2/(G-1)\) over \(50{,}000\) epochs.}\label{fig:wave_results}
\end{figure}

We solve this problem using the Gaussian KAN and the Shepard-normalized Gaussian KAN under the same training setting. 
Figure~\ref{fig:wave_results}(a) shows the validation RMSE versus \(\epsilon\) for \(N=6000\), \(G=20\), and \(W_{\mathrm{BC}}=100\). 
Both models exhibit a clear low-error region near \(\epsilon\approx 10^{-1}\), but the normalized model reaches a lower minimum error. 
Figure~\ref{fig:wave_results}(b) shows the validation RMSE versus epoch at \(\epsilon=2/(G-1)\). 
In this case, the normalized model converges to a smaller error and maintains a more favorable training trajectory. 
Overall, the wave-equation experiment is consistent with the regression and Helmholtz results: Shepard normalization improves accuracy in the practically relevant range of \(\epsilon\).

\section{Conclusion}
\label{sec:conclusion}

We introduced the partition-of-unity Gaussian KAN (PU-GKAN), a kernel-based normalization framework that transfers the classical Shepard partition-of-unity principle to the trainable edge functions of KANs. By normalizing each Gaussian basis over fixed centers, PU-GKAN constructs a partition-of-unity feature map while preserving the standard edge-based architecture and essentially the same trainable coefficient structure as GKAN. The resulting representation reproduces constants exactly at the edge level and requires no subdomains, local subnetworks, or additional trainable parameter blocks.

We formulated GKAN and PU-GKAN from a finite-feature and additive-kernel viewpoint, establishing a direct connection between kernel approximation and KAN architectures. Combined with first-layer conditioning analysis, this formulation supports the practical scale range
\(
\epsilon\in[1/(G-1),\,2/(G-1)]
\).
The upper endpoint often achieves accuracy close to the best result obtained from complete \(\epsilon\)-sweeps, providing an effective reference scale without an expensive parameter search.

The numerical results demonstrate substantial improvements in both accuracy and scale robustness. PU-GKAN reduces validation error by up to \(90\%\), performs consistently across sample-size and center-number variations, and remains effective in higher-dimensional problems. Its benefits also extend to Matérn RBF bases and physics-informed neural networks for Helmholtz and wave equations, showing that the normalization principle is not restricted to Gaussian function approximation. The discontinuous target remains the main case in which the improvement is less uniform.

Overall, PU-GKAN provides a mathematically grounded and broadly applicable normalization framework for kernel-based KANs. It connects Shepard approximation, partition-of-unity methods, and finite-feature kernels with modern edge-based neural architectures while improving the practical reliability of RBF-KANs. Future work may consider adaptive normalized scales, higher-order partition-of-unity constructions, operator learning, and time-dependent PDEs.

\pagebreak

\end{document}